\documentclass[a4paper,11pt]{article}
\usepackage[polish,english]{babel}
\usepackage[utf8]{inputenc}
\usepackage[T1]{fontenc}

\let\babellll\lll
\let\lll\relax
\usepackage{amsmath}
\usepackage{amsfonts}
\usepackage{amssymb} 
\let\lll\babellll 
\usepackage{graphicx}
\usepackage{epstopdf}
\usepackage{color}
\newcommand{\RNumb}{\mathbb{R}}

\newcommand{\ZNumb}{\mathbb{Z}}

\newcommand{\UnitOp}{\hat{1\kern-4.5pt 1}} 
\newcommand{\MatUnit}{1\kern-4.5pt 1} 

\newcommand{\Bra}[1]{\langle #1 \vert}

\newcommand{\Ket}[1]{\vert #1 \rangle}

\newcommand{\BraKet}[2]{\langle #1 \vert #2 \rangle}

\newcommand{\Aver}[1]{\langle #1 \rangle}

\newcommand{\Var}[1]{\mathrm{var}\kern-2pt\left( #1 \right)}
\newcommand{\Norm}[1]{\left\Vert #1 \right\Vert} 
\newcommand{\pev}[1]{\mathrm{pev}\kern-2pt\left( #1 \right)}
\newcommand{\EOp}{\mathsf{E}\kern-1pt\llap{$\vert$}}
\newcommand{\WOp}{\mathsf{W}\kern-1pt\llap{$-$}}
\newcommand{\FOp}{\mathsf{F}\kern-1pt\llap{$\vert$}}

\usepackage{authblk}
\begin{document}

\title{Quantum clock in the projection evolution formalism}

\author{A.~G\'o\'zd\'z 
\thanks{email: Andrzej.Gozdz@umcs.lublin.pl, ORCID: 0000-0003-4489-5136}}
\affil{Institute of Physics \\
Maria Curie--Sk\l{}odowska University in Lublin, Poland}

\author{M.~G\'o\'zd\'z 
\thanks{email: mgozdz@kft.umcs.lublin.pl, ORCID: 0000-0003-4958-8880}}
\affil{Institute of Computer Science \\
Maria Curie--Sk\l{}odowska University in Lublin, Poland}

\maketitle

\abstract{Using the projection evolution (PEv) approach, time can be
  included in the quantum mechanics as an observable. Having the time
  operator, it is possible to explore the temporal structure of various
  quantum events. In the present paper we discuss the possibility of
  constructing a quantum clock, which advances in time during its
  quantum evolution, in each step having some probability to localize
  itself on the time axis in the new position. We propose a working
  two-state model as the simplest example of such a clock.}


\section{Introduction}

Time is one of the most important feature of our physical world. Its
measurement is a fundamental procedure not only for physics. At every
step of technical development there are projects of investigation
leading to construction of better and better clocks which are able to
measure extremaly small time intervals
\cite{zegar2013,zegar2015,zegar2016,zegar2021}.

In the book \cite{muga1} three types of time are discussed: time as a
parameter, dynamical time and time as a quantum observable. The most
consistent with quantum mechanics is the last concept, i.e., time as a
quantum observable considered on the same footing as the other position
operators. In physics, especially in the quantum regime, one requires
more and more precise measurements of time to better understand and
predict evolution of such systems. On the other hand, to measure very
short time intervals one needs to construct a clock which, in fact, is a
specific quantum system working according to quantum rules and, in
addition, is often (weakly) coupled to the measured system.

For many years time was treated in physics as a~universal parameter
enumerating evolution of physical systems. The special and general
relativity change this notion substantially. In quantum mechanics,
however, especially after the publication of the Pauli theorem
\cite{pauli-a,pauli-b} time is still considered as an evolution
parameter.

An overview of the role of time in quantum mechanics can be found, among
others, in \cite{th01,th02,th03,th04,th05,th06,th07,th08,th09,th10,%
  th11,th12,th13,th14,th15,th16,th17,th18,th19,th20,th21,th22,%
  th-pre01,th-pre02,th-pre03,th-pre04,th-pre05,th-pre06,%
  th-rel-mech-a,th-rel-mech-b,th-montevideo,exp-2slits-b,exp-2slits-c,%
  Horwitz2015}.

In this paper we treat time to be a component of a composite quantum
observable of the spacetime position. To be consistent, we use the
projection evolution model (PEv) which recent version can be found in
\cite{PEv2023,PEv2020}. These references present also more extended
introduction to the problem and the corresponding bibliography.

According to the PEv approach quantum time and space positions should be
considered on the same footing. However, one needs to realize that we
have direct access to the space position observable but only indirect
access to the time position observable on the time axis. The latter
requires special measuring devices called clocks. A~proper clock
definition is vital for investigating the structure of the real spacetime.

\section{Projection evolution}

For readers' convenience we summarize the main points of the PEv
approach.

The main assumption of the PEv evolution model is the \textbf{changes
  principle} which states that:

{\it The evolution of a~system is a~random process caused by spontaneous
  changes in the Universe. These spontaneous changes are primary
  processes in the Universe. }

In this approach the changes in the quantum state space happen according
to a probability distribution which is dictated by the properties of the
Universe and its subsystems. A very important attribute of our Universe
is the spacetime itself, which should emerge from the quantum state
space. In this article we consider the flat spacetime.

Every single step of the evolution describes the state of a physical
subsystem (possibly the Universe as a whole) in this state space. It
means that the evolution is not driven by time, which is a part of the
system's description at a given evolution step, but it is driven by an
extra parameter $\tau$. This parameter is not time; it belongs to a
linearly ordered set with no aditional structure required.

In the following we are using $\tau$ which can be represented by
integers $\ZNumb$. This allows to order quantum events and to have the
notion of the predecessor and the successor in any set of physical
events. As a consequence one may expect the existence of a~kind of
pseudo-causality based on the ordering of the quantum events, which
leads to the causality principle in the case of macroscopic physical
systems.  An additional, very important feature of the PEv approach is
that this idea does not need the spacetime as the background, it is
background free.  The spacetime is ``generated'' by the spacetime
position observable.

In the projection evolution formalism we propose to use the generalized
form of the L\"uders \cite{Lueders} type of the projection postulate as
an evolution principle. It allows to reproduce, as a special case, the
standard unitary evolution represented by relativistic and
non-relativistic evolution equations.

In the following we introduce the evolution operators which can
characterize a given physical system. They are formally responsible for
quantum evolution of this object. In this paper we are using the
evolution operators represented by an orthogonal resolution of unity:
\begin{eqnarray}
\label{ProjOpProjections}
&& \EOp(\tau_n;\nu_n)^\dagger=\EOp(\tau_n;\nu_n), \nonumber \\
&& \EOp(\tau_n;\nu_n)\EOp(\tau_n;\nu_n')
   = \delta_{\nu_n\nu_n'} \EOp(\tau_n;\nu_n), \nonumber  \\
&& \sum_{\nu_n \in {\cal Q}_{n}} \EOp(\tau_n;\nu_n)= \UnitOp \, ,
\end{eqnarray}
where $\UnitOp$ denotes the unit operator and $\nu_{n} \in {\cal Q}_{n}$
represents a set of quantum numbers describing quantum states. Different
alternatives representing random choices of quantum states are described
by different sets of quantum numbers $\nu_n$. Such type of operators are
able to describe a wide class of physical systems.

Assume now, that the vector $\Ket{\Phi(\tau_{n-1};\nu_{n-1})}$
represents a given quantum state at the evolution step $\tau_{n-1}$,
where $\nu_{n-1} \in {\cal Q}_{n-1}$.

The changes principle implies that there exists in the Universe
a~physical mechanism, we call it the chooser, which chooses randomly the
next state of the system from the set of states determined by the
projection postulates,
\begin{equation}
\label{eq:NewState}
\Ket{\Phi(\tau_{n};\nu_{n})}= e^{i\alpha_n}
\frac{\EOp(\tau_{n};\nu_{n}) \Ket{\Phi(\tau_{n-1};\nu_{n-1})}}{
\Norm{\EOp(\tau_{n};\nu_{n}) \Phi(\tau_{n-1};\nu_{n-1})} } \, ,
\end{equation}
where the global phase $\alpha_n$ can be chosen arbitrarily and
$\Norm{v}$ denotes the norm of the vector $v$.

In other words, Eq.(\ref{eq:NewState}) determines the set of allowed
states to which a physical system can randomly evolve from the state
$\Ket{\Phi(\tau_{n-1};\nu_{n-1})}$. To fully describe this stochastic
process, one needs to know the probability distribution for getting
a~given state in the next step of the evolution. It is given by:
\begin{equation}
\label{TransProbProjOp}
\pev{\nu_{n-1} \to \nu_{n}}=
\Bra{\Phi(\tau_{n-1};\nu_{n-1})}
\EOp(\tau_{n};\nu_n)
\Ket{\Phi(\tau_{n-1};\nu_{n-1})}\, .
\end{equation}
It is also useful to introduce a tool which facilitates the construction
of the evolution operators in terms of the projection operators. We have
found that required evolution operators can be obtained from some
operators $\WOp$ which we call generators of the projection evolution
\cite{PEv2023}.

\textbf{For a~given evolution step $\tau$ the projection evolution
  generator $\WOp(\tau)$ is defined as a self-adjoint operator which
  spectral decomposition gives the orthogonal resolution of unity
  representing the required set of the evolution operators.}

For example, assuming a discrete spectrum of an evolution generator
$\WOp(\tau_n)$, the spectral theorem gives the following relation
between $\WOp$ and the evolution operators $\EOp$:
$\WOp(\tau_n)= \sum_\nu w_\nu \EOp(\tau_n;\nu) $, where $w_\nu$ are the
eigenvalues of $\WOp$. In the case of a~continuous spectrum one needs to
use the integral form of the spectral theorem.

The evolution generators allow to use the already known quantum
operators to generate the appropriate evolution operators
$\EOp(\tau_n,\nu)$.

\section{Quantum clock in the structureless flat spacetime}

In particle, atomic, molecular and some other branches of physics the
clock can be considered to be independent of the physical system under
consideration. In this case one can treat time as the so-called external
time. However, it is only an approximation because the clock is, in
fact, a part of this system. In models in which time is a quantum
observable, it should be considered as a part of an observable
describing position in spacetime. The position in the quantum spacetime
is, in turn, one of the attributes of physical matter.

In this paper we consider the simplest, approximate model which leads to
the spacetime based on the $L^2(\RNumb^4,d^4x)$ quantum state space,
i.e., the Hilbert space of square integrable functions with the scalar
product (note the integration over time)
$\BraKet{\psi_2}{\psi_1}=\int_{\RNumb^4} d^{4}x \, \psi_2(x)^\star \,
\psi_1(x)$ \cite{PEv2023,PEv2020}.

In a fixed coordinate frame, a~possible realization of the spacetime position
operator can be given by the four-vector operator
$\hat{x}^\mu: \hat{x}^\mu f(x)=x^\mu f(x)$, where $\mu=0,1,2,3$.

We understand the term four-vector as a four-component object which transforms
with respect to a given group of the spacetime transformations. In our case we
think about two groups: either the Galilean or the Lorentz group. It means
that using the same denotations we may consider either a~non-relativistic or
relativistic four dimensional flat spacetime.

The canonically conjugated observable is the four-momentum operator
$\hat{p}_\mu=i\frac{\partial}{\partial x^{\mu}}$. Because time is
a~quantum observable $\hat{t} \equiv \hat{x}^0$, the temporal momentum
$\hat{p}_0$ is the corresponding counterpart. By analogy to the space
components of the momentum operator one can think about $\hat{p}_0$ as
an observable representing the measure of motion in the temporal
dimension. The sign of $\hat{p}_0$ describes the arrow of time. The
temporal momentum can be measured using different equations of motions
which relate it to the spatial and other properties of the system. For
example, the Schr\"odinger equation$(\hat{p}_0-\hat{H})\psi(x,\zeta)=0$,
where the variables $\zeta$ represent some additional degrees of freedom
of the quantum system, relates $\hat{p}_0$ to the Hamiltonian
$\hat{H}$. One can say the same about the Klein-Gordon, Dirac and other
equations of motion.

In principle, we need some kind of ``detectors'' measuring positions in
spacetime. They can be represented by POV measures \cite{Busch1996}. In
the ideal case, this measure is represented by a sharp observable given
by the projection operators
$M_X(x^0,x^1,x^2,x^3)= \Ket{x^0,x^1,x^2,x^3}\Bra{x^0,x^1,x^2,x^3}$,
where the vectors $\Ket{x^0,x^1,x^2,x^3}$ are eigenstates of the
position operators $\hat{x}^\mu$.

In such models we come across the (1+3)-D position measurement. There is no
problem with performing the 3D spatial measurement, however, we do not have
devices which allow to see the whole time axis, i.e., the past, present and
the future. It seems that our material world is, in most cases, rather well
localized in time and that only a narrow temporal window moving along the time
axis is available for us in the experiments. From this perspective a clock has
to be a part of the quantum system, well localized in time, possibly coupled
to other physical subsystems. Having a reference clock, one can construct
other types of clocks as devices synchronized with this clock.  This procedure
allows to introduce the notion of the ``external time'', however only for the
systems which are well decoupled from the reference clock.

We define a~quantum reference clock as a~kind of device, localized on
the time axis and moving with a fixed sign of the average temporal
momentum $\Aver{\hat{p}_0}_{\text{clock state}}$ in spacetime. The sign
is conventional but it defines the arrow of time. The clock localizing
itself in a time interval (temporal window) shows us, usually indirectly
by some kind of an interface (measurement), its position in time. It
means that the clock consists of two subsystems: the proper clock,
evolving from one localization on the time axis to another, and the
interface reading off the clock's time position.

In the PEv model \cite{PEv2020} the average spacetime localization of a
physical system can change only with the change of the evolution
parameter $\tau_n$. In the following we denote by $\tau_n$,
$n=1,2,3,\dots$ the evolution steps of the proper clock, and by
$\tau'_n$, where $\tau_n < \tau'_n < \tau_{n+1}$, the evolution steps at
which the clock interface reads its temporal position. It implies that a
given quantum clock at the evolution step $\tau_n$ is represented by a
set of clock states $\Phi(\tau_n;\nu_n)$. The label $\nu_n$ represents a
set of quantum numbers describing both the temporal and other properties
of observables required for the construction of a quantum clock. For
a~good clock, the variance of the time operator
\begin{equation}
  \hat{t}=\int_\RNumb d x^0 M_T(x^0),
\end{equation}
where
\begin{equation}
  M_T(x^0) = \int_{\RNumb^3} d\mathbf{x} M_X(x^0,\mathbf{x})
\end{equation}
projects onto the subspaces of simultaneous events, should be
small. More precisely, the variance should fulfill the following
inequality:
\begin{equation}
\label{eq:varianceT}
\Var{\hat{t};\Phi(\tau_n;\nu_n)} =
\Bra{\Phi(\tau_n;\nu_n)}
(\hat{t}-\Bra{\Phi(\tau_n;\nu_n)}\hat{t}
\Ket{\Phi(\tau_n;\nu_n)})^2
\Ket{\Phi(\tau_n;\nu_n)} \leq \mathrm{cr} \,  ,
\end{equation}
where $\mathrm{cr}$ is a small number and denotes the clock resolution. The
corresponding expectation value of the time operator within the clock states,
$t=\Bra{\Phi(\tau_n;\nu_n)}\hat{t}\Ket{\Phi(\tau_n;\nu_n)}$, gives the
expected temporal position of the proper clock, i.e., it gives a~parameter $t$
representing the classical time. In the ideal case the clock states are
eigenstates of the time operator with the variance (\ref{eq:varianceT}) being
equal to 0.

In the projection evolution model \cite{PEv2020} an important element is the
construction of the evolution operators for the subsequent steps of the
evolution $\tau_{n-1} \to \tau_{n-1}' \to \tau_{n}$. It means that we have to
describe the following process: for the evolution step $\tau_{n-1}$ the proper
clock is localized at some time $t_{n-1}$, this position is read off by the
interface at $\tau_{n-1}'$ and the proper clock is moving to the next temporal
position at $\tau_{n}$.

In this paper, we propose a set of unitary operations, driven by a~random
variable $\xi \in \RNumb_+$, transforming the evolution generators, not
states, from the previous to the next step of the evolution process:
\begin{equation}
\label{eq:EvolStep}
U(\xi,\gamma)=\exp(i(\xi\hat{p}_0 - \gamma \hat{A})) \, ,
\end{equation}  
where $\gamma$ is a coupling constant. The temporal momentum operator
$\hat{p}_0$ and the self adjoint reconfiguring operator $\hat{A}$
commute, i.e., $[\hat{p}_0,\hat{A}]=0$. The temporal momentum operator
$\hat{p}_0$ is responsible for the motion along the time axis. The
values of the random variable $\xi$ are related to the internal
processes of the clock and to the influence of the environment on the
clock. This leads effectively to a movement of the clock along the time
axis. Direction of the clock motion along the time axis is determined by
the expectation value of the temporal momentum operator $\hat{p}_0$,
calculated in the clock states. The reconfiguring operator $\hat{A}$
changes some internal states of the clock and allows to define the
required clock interface.

Let us denote by $\WOp(\tau_n)$ the evolution generator of the proper clock at
the evolution step $\tau_n$. The unitary transformations (\ref{eq:EvolStep})
form a two-parameter group,
$U(\xi+\xi',(k+k')\gamma)=U(\xi,k\gamma)U(\xi',k'\gamma)$, where
$k,k' \in \ZNumb$. This feature allows to relate the evolution generator
$\WOp(\tau_n)$ to its initial form $\WOp(\tau_0)$:
\begin{equation}
\label{eq:ClockEvGen}
\WOp(\tau_n)=U(\xi_n,\gamma)\,\WOp(\tau_{n-1})\,U(\xi_n,\gamma)^\dagger
=U(\beta_n,n\gamma)\,\WOp(\tau_0)\,U(\beta_n,n\gamma)^\dagger \, ,
\end{equation}
where $\beta_0=0$, $\beta_n=\sum_{k=1}^n \xi_k$ and $\xi_1,\xi_2,\dots \xi_n$
are subsequent values of the random variable $\xi$.

To construct the evolution operators $\EOp(\tau_n;\nu_n)$ from the evolution
generator $\WOp(\tau_n)$ one needs to find its spectral
decomposition. Assuming discrete spectrum of the generator $\WOp(\tau_n)$ one
has to solve the following eigenequation
\begin{equation}
\label{eq:EigenEqEvGen}
\WOp(\tau_n) \Ket{\psi^{(n)}_{bc}} = w^{(n)}_b \Ket{\psi^{(n)}_{bc}} \, , 
\end{equation}
where $c$ represents the possible degeneration of the spectrum. The
transformation property (\ref{eq:ClockEvGen}) implies that it is
sufficient to solve the eigenequation (\ref{eq:EigenEqEvGen}) only for
$\tau_0$ to get all solutions for every $\tau_n$. If
$\Ket{\psi^{(0)}_{bc}} \equiv \Ket{\psi_{bc}}$ are eigenvectors of
$\WOp(\tau_0)$, the states
\begin{equation}
\label{eq:EigSolutionEG}
\Ket{\psi^{(n)}_{bc}}= U(\beta_n,n\gamma) \Ket{\psi_{bc}} 
\end{equation}
are eigenstates of $\WOp(\tau_n)$. The eigenvalues
$w^{(n)}_b = w^{(0)}_b \equiv w_b$ are independent of $n$.

The distribution of $\xi$ should have a~pronounced maximum very close to
zero\footnote{In this work we assume $\xi>0$ but the general condition
  is, that $\xi\in\RNumb$ has a~distribution around zero, assymetric
  towards $\xi>0$ values. This will result in the net time flow in the
  positive direction of the time axis. We will discuss this problem in
  a~subsequent paper.}. It implies, that the next instant of classical
time during the evolution should be very close to the previous one. The
probability distribution for $\xi$ is a parameter of the clock. It seems
that this distribution is strongly related to both the construction of
the clock, and the structure of spacetime.

The second component of the clock is an interface allowing to read the
clock. The interface is therefore a measuring device and its full evolution
operator has to contain the appropriate projection operators.

The projection evolution generators simplify the construction of the required
evolution operators for a quantum reference clock. This method will be used in
the next section.

\subsection{The proper clock}

For a given observer the quantum spacetime splits into time and 3D position
space. In the following, to present the clock idea we choose a non-covariant
description of a proper clock and its interface.

Because we treat time on the same footing as the other coordinates one
expects that every interaction $V_{int}(x-y)$, where
$x=(t_x,\mathbf{x}),y=(t_y,\mathbf{y})$ are spacetime points, depends
not only on the positions in space but also on positions in time
\cite{PEv2023}. One can say the same about the effective potentials
$V(t,\mathbf{x})$. According to observations, in a wide range of its
density, the physical matter is well localized in time.  This supports
the parametric time as a good approximation. On the other hand, what is
even more important, this feature also suggests that the interaction in
the temporal dimension has a very short range. In the normal density
matter the temporal parts of interactions among many particles lead to
a~temporal mean-field effective potential operator $V_T(\hat{t})$. One
can find similar effect in nuclear physics, where the mean-field
approach to a short range nuclear interaction is a~good approximation.

In the following, we present a nonrelativistic proper clock described by
a modified Schr\"odinger type of quantum motion in the structureless
flat spacetime \cite{PEv2023}. Relativistic versions of the clock can be
constructed in a similar way.

Following the discussion about the evolution generators described in
\cite{PEv2023}, the initial projection evolution generator
$\WOp(\tau_0)$ for a proper clock is postulated in the following form:
\begin{equation}
\label{eq:0PEvGenNonrelC}
\WOp(\tau_0) = 
\hat{p}_0 - \hat{H} - \frac{\hat{p}_0^2}{2m_T} - V_T(\hat{t}),
\end{equation}
where the temporal momentum operator is
$\hat{p}_0=i\frac{\partial}{\partial t}$, the Hamiltonian $\hat{H}$ acts on
functions of the position variables $\mathbf{x} \in \RNumb^3$ and potentially
on some intrinsic variables, i.e., it does not depend on time. The signs in
front of the temporal part of the generator \eqref{eq:0PEvGenNonrelC} are
chosen to keep the coefficient $m_T$ positive. By analogy to the spatial
kinetic term in the Hamiltonian $\hat{H}$ the term $m_T$ can be called ``the
temporal inertia''.

The proper clock is determined by the generator, which describes the physical
system localized in time, and the Hamiltonian $\hat{H}$ which allows to build an
appropriate interface in its state space. The proper clock is localized on the
time axis by the effective temporal potential operator $ V_T(\hat{t})$.

Using the unitary operator (\ref{eq:EvolStep}) one gets the evolution generator
for any arbitrary evolution step $\tau_n$:
\begin{equation}
\label{eq:3PEvGenNonrelC}  
\WOp(\tau_n) = U(\beta_n,n\gamma)\,\WOp(\tau_0)\,U(\beta_n,n\gamma)^\dagger
= \hat{p}_0 - \hat{H}^{(n)} - \frac{\hat{p}_0^2}{2m_T} - V^{(n)}_T(\hat{t}) \, ,
\end{equation}
where the potential localizing the clock in time is given by
\begin{equation}
\label{eq:LocPotT}
  V_T^{(n)}(\hat{t}) = V_T \left( \hat{t} - \beta_n \UnitOp \right) \, . 
\end{equation}
The modified Hamiltonian has the following form:
\begin{equation}
\label{eq:TansfNham}
\hat{H}^{(n)} = 
\exp\left(-i n\gamma\, \hat{A} \right) 
\hat{H} 
\exp\left(+i n\gamma\,\hat{A} \right) \, . 
\end{equation}
According to the definition of the projection evolution generators we have to
construct the spectral decomposition of $\WOp(\tau_n)$. For this purpose one
needs to look for the eigenfunctions of $\WOp(\tau_0)$. In our case we can
find them by separating the variables,
\begin{equation}
\label{eq:psi}
\psi_{\lambda\kappa\nu\mu}(t,\mathbf{x})
= \chi_{\lambda\kappa}(t) \phi_{\nu\mu}(\mathbf{x}) \, , 
\end{equation}
where $\chi_{\lambda\kappa}(t) := e^{-i m_T t}
f_{\lambda\kappa}(t)$. The indices $\kappa$ and $\mu$ indicate a
possible degeneration of the solutions.  In the following we assume that
the temporal functions $\chi_{\lambda\kappa}(t)$ represent the states
which, on average, move in the positive direction of time. This means
that the temporal functions $\chi_{\lambda\kappa}(t)$ are the states for
which the expectation value of the temporal momentum operator,
\begin{equation}
\Bra{\chi_{\lambda\kappa}} \hat{p}_0 \Ket{\chi_{\lambda\kappa}}
= m_T
+ \int_{R} dt\, f_{\lambda\kappa}(t)^\star \hat{p}_0 f_{\lambda\kappa}(t) ,
\end{equation}
is a positive number.

The functions $\phi_{\nu\mu}(\mathbf{x})$ denote the eigenvectors of the
Hamiltonian $\hat{H}$,
\begin{equation}
\label{eq:HamEigenEq}
\hat{H} \phi_{\nu\mu}(\mathbf{x}) = E_\nu \phi_{\nu\mu}(\mathbf{x}) \, ,
\end{equation}
where $E_\nu$ are the eigenvalues of $\hat{H}$.

The function $f_{\lambda\kappa}(t)$ solves the equation
\begin{equation}
\label{eq:fFun}
\left( \frac{\hat{p}_0^2}{2m_T} + V_T(\hat{t}) \right)
f_{\lambda\kappa}(t)
= \left( \frac{m_T}{2} - \epsilon_\lambda^{(T)} \right)
f_{\lambda\kappa}(t),
\end{equation}
while together with the exponential function it constitutes the full
solution of the equation
\begin{equation}
\label{eq:EigenEqT}
\left( \hat{p}_0 - \frac{\hat{p}_0^2}{2m_T} - V_T(\hat{t}) \right)
\chi_{\lambda\kappa}(t)
 = \epsilon_\lambda^{(T)}\chi_{\lambda\kappa}(t) \, .
\end{equation}
This allows one to write the eigenequation for $\WOp(\tau_0)$ as
\begin{equation}
\label{eq:EigenEqW0}
\WOp(\tau_0) \psi_{\lambda\kappa\nu\mu}(t,\mathbf{x}) =
w_{b(\lambda,\nu)} \psi_{\lambda\kappa\nu\mu}(t,\mathbf{x}),
\end{equation}
with $\psi_{\lambda\kappa\nu\mu}$ given by Eq.~(\ref{eq:psi}) and
\begin{equation}
\label{eq:EigenValw}
w_{b(\lambda,\nu)} = \epsilon^{(T)}_\lambda - E_{\nu}  \, ,
\end{equation}
where the function labelling the eigenvalues of $\WOp(\tau_0)$ fulfil
the condition:
$b(\lambda',\nu')=b(\lambda,\nu) \Leftrightarrow
w_{b(\lambda',\nu')}=w_{b(\lambda,\nu)}$.

Collecting the above partial solutions the eigenstates of $\WOp(\tau_n)$
for a given evolution step $\tau_n$ are given by the vectors
\begin{equation}
\label{eq:EigenFunWn}
\psi^{(n)}_{\lambda\kappa\nu\mu}(t,\mathbf{x})
= U(\beta_n,n\gamma) \psi_{\lambda\kappa\nu\mu}(t,\mathbf{x})
= \chi^{(n)}_{\lambda\kappa}(t) \phi^{(n)}_{\nu\mu}(\mathbf{x})  \, ,
\end{equation}
where $\chi^{(n)}_{\lambda\kappa}(t):= \chi_{\lambda\kappa}(t-\beta_n)$ and
$\phi^{(n)}_{\nu\mu}(\mathbf{x}):=\exp(-i n\gamma \hat{A})
\phi_{\nu\mu}(\mathbf{x})$. 

The evolution operators generated by $\WOp(\tau_n)$ are the projections onto
eigenspaces of the evolution generator,
\begin{equation}
\label{eq:EvolGenClock}
\EOp(\tau_n;w) = \sum_{\lambda\nu} \delta(w=\epsilon^{(T)}_{\lambda} -E_{\nu})
\sum_{\kappa\mu}
\Ket{\psi^{(n)}_{\lambda\kappa\nu\mu}}
\Bra{\psi^{(n)}_{\lambda\kappa\nu\mu}}
= U(\beta_n,n\gamma) \EOp(\tau_0;w) U(\beta_n,n\gamma)^\dagger \, .
\end{equation}
The Kronecker's type function is defined as:
$\delta(w=\epsilon^{(T)}_{\lambda} -E_{\nu})=1$ if
$w=\epsilon^{(T)}_{\lambda} -E_{\nu}$, otherwise it is equal to 0. The sums will
be substituted by some integrals in the case of a~continuous spectrum.

\subsection{The clock interface}

To read the change of the internal state of the proper clock, we
introduce the interface evolution operators denoted by
$\EOp_I(\tau_{n}',\nu)$, where $\tau_{n} < \tau_{n}' <\tau_{n+1}$
describes an intermediate event between $\tau_{n}$ and
$\tau_{n+1}$. They form an orthogonal resolution of unity projecting
onto eigenstates of the Hamiltonian $\hat{H}$ showing in which
eigenstate of $\hat{H}$ the clock actually is. Changes of the states of
the Hamiltonian along the time axis represent the clock ticks.  The
clock interface should disturb neither the proper clock localization in
time nor its movement along the time axis but it should show in which
eigenstate of the Hamiltonian the proper clock is.

Let us denote an orthonormal basis
$g_{s\nu\mu}(t,x)=e_{s}(t) \phi_{\nu\mu}(\mathbf{x})$ in the state
space. We obtain the required properties by assuming that
\begin{equation}
\label{eq:ClockInterface}
\EOp_I(\tau_{n}',\nu)= \sum_{\mu}
\Ket{\phi_{\nu\mu}}
\left( \sum_s \Ket{e_{s}} \Bra{e_{s}} \right)
\Bra{\phi_{\nu\mu}} \equiv \EOp_I(\nu) \, ,
\end{equation}
where $\left( \sum_s \Ket{e_{s}} \Bra{e_{s}} \right)=\UnitOp_T$ is the unit
operator in the time domain. The projection operators
(\ref{eq:ClockInterface}) are independent of $\tau_{n}'$, i.e., we keep the
same interface for every evolution step.

\subsection{The clock}

The evolution operators described above allow to construct different
reference clocks using different forms of the operators:
$\hat{A},\hat{H}$ and $V_T(\hat{t})$.

The clock starts from any arbitrary quantum state $\Ket{\Phi_0}$. The evolution
operator $\EOp(\tau_0;u_0)$ prepares the initial state of the clock
\begin{equation}
\label{eq:Tau0State}
\Phi(\tau_0;u_0;t,\mathbf{x})
=\frac{\EOp(\tau_0;u_0)\Phi_0(t,\mathbf{x})}{\Norm{\EOp(\tau_0;u_0)\Phi_0}} \, ,
\end{equation}
where $u_0$ represents, according to the PEv formalism, a randomly chosen
eigenvalue $w_{b(\lambda,\nu)}$  of the evolution generator $\WOp(\tau_0)$.
  
The state (\ref{eq:Tau0State}) is read by the interface
\begin{equation}
\label{eq:Tau0PrimState}
\Phi(\tau_0';\sigma_0,u_0;t,\mathbf{x)}
=\frac{\EOp_I(\sigma_0) \Phi(\tau_0;u_0;t,\mathbf{x})}{
  \Norm{\EOp_I(\sigma_0) \Phi(\tau_0;u_0)}} \, ,
\end{equation}
where $\sigma_0$ represents a randomly chosen eigenvalue $E_{\sigma_0}$ of the
Hamiltonian $\hat{H}$.
The simplest choice of the initial state is any eigenstate of the evolution
generator (\ref{eq:psi}),
\begin{equation}
\label{eq:Tau0PrimSimplest}
\Phi_0(t,\mathbf{x}) = \psi_{\lambda_0\kappa_0\nu_0\mu_0}(t,\mathbf{x})
\equiv
\chi_{\lambda_0\kappa_0}(t) \phi_{\nu_0\mu_0}(\mathbf{x}) \, .
\end{equation}
Then $\Phi(\tau_0';\sigma_0,u_0;t,\mathbf{x})=\Phi_0(t,\mathbf{x})$, where
$\sigma_0=\nu_0$ and $u_0=\epsilon^{T}_{\lambda_0} -E_{\nu_0}$.

To simplify the notation let us denote by $(\sigma,u)_{n}$ the sequence
of quantum numbers which gives a possible evolution path
$$
(\sigma,u)_{n} = (\sigma_{n},u_{n},\sigma_{n-1},u_{n-1},\dots,\sigma_{0},u_{0})
= (\sigma_n,u_n,(\sigma,u)_{n-1}).
$$

The subsequent cycles of the clock are described by the following
recurrence relations 
\begin{equation}
\label{eq:TauNState}
\Phi(\tau_n;u_n,(\sigma,u)_{n-1};t,x)
=\frac{\EOp(\tau_n;u_n) \Phi(\tau_{n-1}';(\sigma,u)_{n-1};t,x)}{
\Norm{\EOp(\tau_n;u_n) \Phi(\tau_{n-1}';(\sigma,u)_{n-1})} } \, ,
\end{equation}
and
\begin{eqnarray}
\label{eq:TauNPrimState}
\Phi(\tau_n';(\sigma,u)_{n};t,x)
&=&\frac{\EOp_I(\sigma_n) \Phi(\tau_{n};u_n,(\sigma,u)_{n-1};t,x)}{
  \Norm{ \EOp_I(\sigma_n) \Phi(\tau_{n};u_n,(\sigma,u)_{n-1})} }
\nonumber \\
&=& \frac{\EOp_I(\sigma_n) \EOp(\tau_n;u_n)
  \Phi(\tau_{n-1}';(\sigma,u)_{n-1};t,x)}{
  \Norm{\EOp_I(\sigma_n) \EOp(\tau_n;u_n)
  \Phi(\tau_{n-1}';(\sigma,u)_{n-1})   } } \, ,
\end{eqnarray}
where the probabilities of choosing next states are given by
(\ref{TransProbProjOp}), i.e., by the denominators of
(\ref{eq:TauNState}) and (\ref{eq:TauNPrimState}), respectively.

The action of the evolution operator $\EOp_I(\sigma_n)\EOp(\tau_n;u_n)$
on any function $\Phi(\tau_{n-1}';(\sigma,u)_{n-1};t,x)$ can be written
as
\begin{equation}
\label{eq:EOpIEOp}
  \EOp_I(\sigma_n)\EOp(\tau_n;u_n) \Phi(\tau_{n-1}';(\sigma,u)_{n-1};t,x)
  = \mathcal{G}((\sigma,u)_{n};\lambda\kappa\mu')
  \chi^{(n)}_{\lambda\kappa}(t) \phi_{\sigma_n\mu'}(\mathbf{x}) \, ,
\end{equation}
where the expansion coefficients, taking into account
(\ref{eq:EvolGenClock}) and (\ref{eq:ClockInterface}), read
\begin{equation}
\label{eq:EOpIEOpG}
\mathcal{G}((\sigma,u)_{n};\lambda\kappa\mu')
= \sum_{\nu\mu} \delta(u_n=\epsilon^{(T)}_{\lambda} -E_{\nu})\,
\BraKet{\phi_{\sigma_{n}\mu'}}{\phi^{(n)}_{\nu\mu}} \,
\BraKet{\psi^{(n)}_{\lambda\kappa\nu\mu}}{ \Phi(\tau_{n-1}';(\sigma,u)_{n-1})}
\, . 
\end{equation}
Because the functions
$\chi^{(n)}_{\lambda\kappa}(t) \phi_{\sigma_n\mu'}(\mathbf{x})$ and
$\chi^{(n)}_{\lambda'\kappa'}(t) \phi_{\sigma_n'\mu''}(\mathbf{x})$ are
orthonormal, the coefficient $N(\tau_n;(\sigma,u)_{n})$ which normalizes 
Eq.~(\ref{eq:EOpIEOp}) can be expressed by the coefficients
(\ref{eq:EOpIEOpG})
\begin{equation}
\label{eq:EOpIEOpNorm}
N(\tau_n;(\sigma,u)_{n})^2
= \Norm{\EOp_I(\sigma_n) \EOp(\tau_n;u_n) \Phi(\tau_{n-1}';(\sigma,u)_{n-1}) }^2
= \sum_{\lambda\kappa\mu'}
|\mathcal{G}((\sigma,u)_{n};\lambda\kappa\mu')|^2 \, .
\end{equation}
Using the coefficients $\mathcal{G}$ the clock state \eqref{eq:TauNPrimState}
can be rewriten as
\begin{equation}
\label{eq:TauNPrimState2}
\Phi(\tau_n';(\sigma,u)_{n};t,x)
= N(\tau_n;(\sigma,u)_{n})^{-1}
\sum_{\lambda\kappa\mu'}
\mathcal{G}((\sigma,u)_{n};\lambda\kappa\mu')
\chi^{(n)}_{\lambda\kappa}(t) \phi_{\sigma_n\mu'}(\mathbf{x}) \, .
\end{equation}
The definition \eqref{eq:EOpIEOpG} and Eqs.~\eqref{eq:EOpIEOpNorm} and
\eqref{eq:TauNPrimState2} lead to the recurrence relation for the
$\mathcal{G}$-coefficients
\begin{eqnarray}
\label{eq:EOpIEOpG2}
&&\mathcal{G}((\sigma,u)_{n};\lambda\kappa\mu')
= N(\tau_{n-1};(\sigma,u)_{n-1})
\sum_{\nu_1\mu_1} \delta(u_n=\epsilon^{(T)}_{\lambda}-E_{\nu_1})
\BraKet{\phi_{\sigma_{n} \mu'}}{\phi^{(n)}_{\nu_1 \mu_1}} \nonumber \\
&& \sum_{\lambda_2 \kappa_2 \mu_2'}
\mathcal{G}((\sigma,u)_{n-1};\lambda_2 \kappa_2\mu_2')
\BraKet{\chi^{(n)}_{\lambda\kappa}}{\chi^{(n-1)}_{\lambda_2\kappa_2}}
\BraKet{\phi^{(n)}_{\nu_1\mu_1}}{\phi_{\sigma_{n-1}\mu_2'}}.
\end{eqnarray}
Having the $\mathcal{G}$-coefficients, one can calculate the transition
probability between the subsequent readings off the interface. The square of
the normalization factor represents this probability:
\begin{eqnarray}
\label{eq:TransProbSubseqStates}
&& \pev{\tau_{n-1}'\to \tau_{n} \to  \tau_{n}'} \nonumber \\
&& =|\BraKet{\Phi(\tau_{n}';(\sigma,u)_{n})}{
     \Phi(\tau_{n};u_n,(\sigma,u)_{n-1})}
   \BraKet{\Phi(\tau_{n};u_n,(\sigma,u)_{n-1})}{
     \Phi(\tau_{n-1}';(\sigma,u)_{n-1})} |^2 \nonumber \\
&& =\Norm{\EOp_I(\sigma_n) \EOp(\tau_n;u_n)
  \Phi(\tau_{n-1}';(\sigma,u)_{n-1})}^2
= N(\tau_n;(\sigma,u)_{n})^2 \, .
\end{eqnarray}
In our model, the clock shifts subsequently on the time axis about the random
variable $\xi$. However, our knowledge about this shift is supplied only by
the clock interferface. It is important to calculate the expectation value of
the time operator within the states corresponding to reading off the
interface.  For this purpose, let us denote by $t_n$ the average value of the
time operator in the state \eqref{eq:TauNPrimState}, i.e.,
$t_n=\Bra{\Phi(\tau_{n}';(\sigma,u)_{n})}\hat{t}
\Ket{\Phi(\tau_{n}';(\sigma,u)_{n})}$.  Using \eqref{eq:TauNPrimState} and the
following matrix elements:
\begin{eqnarray}
\label{eq:MatElemChi}
\Bra{\chi^{(n)}_{\lambda_1\kappa_1}}
\hat{t}
\Ket{\chi^{(n)}_{\lambda_2\kappa_2}}
 &=&\int_{\RNumb} dt\,
 f_{\lambda_1\kappa_1}(t-\beta_{n})^\star t f_{\lambda_2\kappa_2}(t-\beta_n)
\nonumber \\ 
 &=&\delta_{\lambda_1\lambda_2}\delta_{\kappa_1\kappa_2}\beta_{n}
 + \int_{\RNumb} dt\,
 f_{\lambda_1\kappa_1}(t)^\star\, t\, f_{\lambda_2\kappa_2}(t) \nonumber \\
  &=&\delta_{\lambda_1\lambda_2}\delta_{\kappa_1\kappa_2}\beta_{n}
 + \Bra{\chi_{\lambda_1\kappa_1}} \hat{t} \Ket{\chi_{\lambda_2\kappa_2}},
\end{eqnarray}
the required expectation value reads
\begin{eqnarray}
\label{eq:ExpValTRead}
&& t_n = \beta_n + \left(\frac{1}{N(\tau_n;(\sigma,u)_{n})}\right)^2
\Bigg[ \sum_{\lambda_1\kappa_1} \sum_{\lambda_2\kappa_2}
\nonumber \\
&& \sum_{\mu'} \mathcal{G}((\sigma,u)_n;\lambda_1\kappa_1,\mu')^\star
\mathcal{G}((\sigma,u)_n;\lambda_2\kappa_2,\mu')
\Bigg]
\Bra{\chi_{\lambda_1\kappa_1}} \hat{t} \Ket{\chi_{\lambda_2\kappa_2}} \, .
\end{eqnarray}
The ideal clock should show the value $t_n=\beta_n$, i.e., the place where the
clock is localized on the time axis. The interface, however, is also a quantum
device and acts randomly. There is therefore a finite probability that the
proper clock moves to the next position on the time axis, but the interface
does not change its state, i.e., the ``hand of the clock'' does not advance
forward. This implies that for a good clock the second term in
\eqref{eq:ExpValTRead} should be always close to zero.

\section{A~two-state clock}

As an example, we build a schematic clock for which there is no summation over
$\lambda,\mu,\kappa$ and $\nu$ in Eq.~(\ref{eq:EvolGenClock}).

This is often the case when both (\ref{eq:HamEigenEq}) and
(\ref{eq:fFun}) have discrete nondegenerate spectra, which allows to
rewrite the evolution operators (\ref{eq:EvolGenClock}) in a simpler
form:
\begin{equation}
\label{eq:2EvolGenClock}
\EOp(\tau_n;\lambda\nu)
=\Ket{\psi^{(n)}_{\lambda\nu}}\Bra{\psi^{(n)}_{\lambda\nu}}.
\end{equation}
We express the clock Hamiltonian describing the clock interface by making use
of the spectral theorem,
\begin{equation}
\label{eq:ClockHamiltonian}
\hat{H}=\sum_{\nu=0}^\infty E_\nu \Ket{\phi_\nu}\Bra{\phi_\nu} \, .
\end{equation}
The reconfiguration operator is taken in the following form:
\begin{equation}
\label{eq:RecOpA}
\hat{A}=\Ket{\phi_0}\Bra{\phi_1} + \Ket{\phi_1}\Bra{\phi_0} \, .
\end{equation}
This means that our clock interface is oscillating between two states,
$\Ket{\phi_0}$ and $\Ket{\phi_1}$, and the corresponding evolution
operators, reading the actual state of the interface, are given by
\begin{equation}
\label{eq:EvolOpReadInt}
\EOp_I(\sigma)=\Ket{\phi_\sigma}\Bra{\phi_\sigma},\quad  \sigma=0,1.
\end{equation}
The state of the proper clock and its interface at the evolution step
$\tau'_{n-1}$ is $\Ket{\Phi(\tau'_{n-1};(\sigma,\lambda,\nu)_{n-1}
  )}$. To get to the next step of the evolution one needs to calculate
\begin{equation}
  \EOp(\tau_n;\lambda_n,\nu_n)
  \Ket{\Phi(\tau'_{n-1};(\sigma,\lambda,\nu)_{n-1} )}
  = \BraKet{\psi^{(n)}_{\lambda_n\nu_n}}{
    \Phi(\tau'_{n-1};(\sigma,\lambda,\nu)_{n-1} )}
  \Ket{\psi^{(n)}_{\lambda_n\nu_n}}
  \label{eq:TauNStateSimple}
\end{equation}
and
\begin{eqnarray}
  && \EOp_I(\sigma_n) \EOp(\tau_n;\lambda_n,\nu_n)
  \Ket{\Phi(\tau'_{n-1};(\sigma,\lambda,\nu)_{n-1} )}
     = \nonumber \\
  && \BraKet{\phi_{\sigma_n}}{\phi^{(n)}_{\nu_n}}
  \BraKet{\psi^{(n)}_{\lambda_n\nu_n}}{
    \Phi(\tau'_{n-1};(\sigma,\lambda,\nu)_{n-1} )}
  \Ket{\chi^{(n)}_{\lambda_n}} \Ket{\phi_{\sigma_n}} \, .
\label{eq:TauNPrimStateSimple} 
\end{eqnarray}
Because in the formulas \eqref{eq:TauNState} and
\eqref{eq:TauNPrimState} the common multiplication factors in the
nominators and denominators can be simplified, and overall phases are
unimportant, the resulting clock states, after simplification of
notation, are represented by the vectors
\begin{eqnarray} 
&&\Ket{\Phi(\tau_{n};(\sigma,\lambda,\nu)_{n} )} 
= \Ket{\Phi(\tau_{n};\lambda_n \nu_n)}
\equiv \Ket{\psi^{(n)}_{\lambda_n\nu_n}},
 \label{eq:TauNStateSimple2} \\
 && \Ket{\Phi(\tau'_{n};(\sigma,\lambda,\nu)_{n})}
= \Ket{\Phi(\tau'_{n};\sigma_{n},\lambda_{n})} 
\equiv \Ket{\chi^{(n)}_{\lambda_n}} \Ket{\phi_{\sigma_n}} \, .
\label{eq:TauNPrimStateSimple2} 
\end{eqnarray}
The clock states \eqref{eq:TauNStateSimple2} and
\eqref{eq:TauNPrimStateSimple2} allow to calculate the transition
probability between the states representing two subsequent readings from
the clock interface. The probability of passing the evolution path
$$
(\tau',\sigma,\lambda,\nu)_{n-1} \to
(\tau_{n},\lambda_{n},\nu_{n},(\tau,\sigma,\lambda,\nu)_{n-1}) \to
(\tau',\sigma,\lambda,\nu)_{n}
$$
according to the expression \eqref{eq:TransProbSubseqStates} reads
\begin{equation}
\label{eq:TransProbInterf}
\mathrm{pev}'=|\BraKet{\Phi(\tau'_{n};\sigma_n,\lambda_n)}{
  \Phi(\tau_{n};\lambda_n,\nu_n)}  
\BraKet{\Phi(\tau_{n};\lambda_n,\nu_n)}{
\Phi(\tau'_{n-1};\sigma_{n-1},\lambda_{n-1})}|^2 \, .
\end{equation}
Using relations (\ref{eq:TauNStateSimple}) and
(\ref{eq:TauNPrimStateSimple}), one can cast (\ref{eq:TransProbInterf})
in the form
\begin{equation}
  \label{eq:eq:TransProbInterf2}
  \mathrm{pev}'=
  |\BraKet{\chi_{\lambda_n}^{(n)}}{\chi_{\lambda_{n-1}}^{(n-1)}}|^2
  |\BraKet{\phi_{\sigma_n}}{\phi_{\nu_n}^{(n)}}
  \BraKet{\phi_{\nu_n}^{(n)}}{\phi_{\sigma_{n-1}}}|^2.
\end{equation}
For the reconfiguration operator (\ref{eq:RecOpA}), the
$\hat{A}$-dependent part of the $U(\xi,\gamma)$ operator
(\ref{eq:EvolStep}) is given by
\begin{equation}
  \exp(-i \gamma \hat{A}) = \hat{1} -i \sin\gamma\hat{A}
  + \left(\cos\gamma -1\right) \hat{A}^2,
\end{equation}
where $\hat{1}$ is the unit operator in the full state space, while
$\hat{A}$ acts in its two-dimensional subspace only. It follows that the
scalar product reads
\begin{equation}
  \BraKet{\phi_{\sigma_n}}{\phi_{\nu_n}^{(n)}} =
  \Bra{\phi_{\sigma_n}} \exp(-i \gamma \hat{A}) \Ket{\phi_{\nu_n}} =
  \delta_{\sigma_n\nu_n} \cos\gamma
  -i (\delta_{\sigma_n 1} \delta_{\nu_n 0} +
  \delta_{\sigma_n 0} \delta_{\nu_n 1}) \sin\gamma,
\end{equation}
which allows to calculate the transition amplitudes:
\begin{eqnarray}
  && \BraKet{\phi_1}{\phi_1^{(n)}}\BraKet{\phi_1^{(n)}}{\phi_0} =
     \frac{i}{2}\sin(2\gamma), \label{eq:amp1} \\
  && \BraKet{\phi_1}{\phi_0^{(n)}}\BraKet{\phi_0^{(n)}}{\phi_0} =
     -\frac{i}{2}\sin(2\gamma), \\
  && \BraKet{\phi_0}{\phi_1^{(n)}}\BraKet{\phi_1^{(n)}}{\phi_0} =
     \sin^2\gamma, \\
  && \BraKet{\phi_0}{\phi_0^{(n)}}\BraKet{\phi_0^{(n)}}{\phi_0} =
     \cos^2\gamma, \\
  && \BraKet{\phi_0}{\phi_1^{(n)}}\BraKet{\phi_1^{(n)}}{\phi_1} =
     -\frac{i}{2}\sin(2\gamma), \\
  && \BraKet{\phi_0}{\phi_0^{(n)}}\BraKet{\phi_0^{(n)}}{\phi_1} =
     \frac{i}{2}\sin(2\gamma), \\
  && \BraKet{\phi_1}{\phi_1^{(n)}}\BraKet{\phi_1^{(n)}}{\phi_1} =
     \cos^2\gamma, \\
  && \BraKet{\phi_1}{\phi_0^{(n)}}\BraKet{\phi_0^{(n)}}{\phi_1} =
     \sin^2\gamma.  \label{eq:amp8}
\end{eqnarray}
To register the time passing is equivalent for the clock to change the state
from $\sigma_n=1$ to $\sigma_n=0$ or from $0$ to $1$, i.e., the clock
``clicks''. Using Eq.~\eqref{eq:eq:TransProbInterf2} and the appropriate
amplitudes \eqref{eq:amp1}--\eqref{eq:amp8} gives the probabilities that
the clock ``clicks'' in the form
\begin{eqnarray}
  \label{eq:ProbTick}
&& p(\lambda_{n}):= \mathrm{Prob}(\sigma_{n-1}=1 \to \sigma_n=0) =
  \mathrm{Prob}(\sigma_{n-1}=0 \to \sigma_n=1) \nonumber \\
&& = \frac{1}{2}\sin^2(2\gamma) 
  |\BraKet{\chi_{\lambda_n}^{(n)}}{\chi_{\lambda_{n-1}}^{(n-1)}}|^2
  \le \frac{1}{2} \, .
\end{eqnarray}
The probability of the opposite event, i.e., that the clock changes its
state without the ``click'' is $1- p(\lambda_{n})$.
  
The probability that the clock ``clicks'' exactly in the $\ell$-th step is
given by
\begin{equation}
  \label{eq:prob_clickL}
  p(\lambda_{\ell}) \prod_{k=1}^{\ell-1} (1-p(\lambda_{k}))  \, .
\end{equation}
The probability amplitude
$\BraKet{\chi_{\lambda_n}^{(n)}}{\chi_{\lambda_{n-1}}^{(n-1)}}$ can be
expressed as
\begin{equation}
\label{eq:ChiAmp}
\BraKet{\chi_{\lambda_n}^{(n)}}{\chi_{\lambda_{n-1}}^{(n-1)}}
=e^{-im_T \xi_n} \int_{R} dt\, f_{\lambda_n}(t-\xi_n)^\star
f_{\lambda_{n-1}}(t) \, .
\end{equation}
The elementary step of movement along the time axis should be extremely small,
i.e., the expectation value of $\Aver{\xi_k} \approx 0 $, for every step $k$
of the clock evolution. This implies that
$|\BraKet{\chi_{\lambda_n}^{(n)}}{\chi_{\lambda_{n-1}}^{(n-1)}}|^2=1$ with a
very good approximation and one can write
$p(\lambda_{\ell})\approx p=\frac{1}{2}\sin^2(2\gamma)$.

Let us calculate the average value of the time operator $\hat{t}$. It is
given by
\begin{equation}
  \Bra{\Phi(\tau'_{n};\sigma_n,\lambda_n)} \hat{t}
  \Ket{\Phi(\tau'_{n};\sigma_n,\lambda_n)} =
  \beta_n + \Bra{\chi_\lambda} \hat{t} \Ket{\chi_\lambda}.
\end{equation}
Since $\chi_\lambda = e^{-i~m_T t} f_\lambda(t)$, we have
\begin{equation}
  \label{eq:chiTchi}
  \Bra{\chi_\lambda} \hat{t} \Ket{\chi_\lambda} =
  \int_\RNumb dt \ t |f_\lambda(t)|^2.
\end{equation}
A~good clock is characterized by the term (\ref{eq:chiTchi}) being as
small as possible. If $f_\lambda(-t) = e^{i\alpha} f_\lambda(t)$ this
term vanishes. It follows, that after $n$ steps of evolution, the
localization of the clock on the time axis is close to $\beta_n$.

In a~similar way one may obtain the average value of the temporal
momentum operator,
\begin{equation}
  \label{eq:phiPphi}
  \Bra{\Phi(\tau'_{n};\sigma_n,\lambda_n)} \hat{p}_0
  \Ket{\Phi(\tau'_{n};\sigma_n,\lambda_n)} =
  m_T + \int_\RNumb dt f_\lambda(t)^\star \hat{p_0} f_\lambda(t).
\end{equation}
Since $\hat{p}_0=i\frac{\partial}{\partial t}$, the condition
$f_\lambda(-t) = e^{i\alpha} f_\lambda(t)$ implies that the second term
in Eq.~(\ref{eq:phiPphi}) vanishes, which results in a~constant arrow of time:
\begin{equation}
  \Bra{\Phi(\tau'_{n};\sigma_n,\lambda_n)} \hat{p}_0
  \Ket{\Phi(\tau'_{n};\sigma_n,\lambda_n)} = m_T > 0.
\end{equation}

To estimate how long, on average, do we have to wait until a~``click''
appears, we calculate the variance of $\ell$. The normalized probability that
a~single ``click'' takes place at any of the steps, during an
$\ell$-step evolution, is given by
\begin{equation}
  \label{eq:Prob}
  \mathrm{Prob}(\mathrm{click},\ell) =
  \frac{\ell p(1-p)^{\ell-1}}{\sum_{\ell=1}^\infty \ell p(1-p)^{\ell-1}} =
  \ell p^2 (1-p)^{\ell-1}.
\end{equation}
This results in the average $\ell$,
\begin{equation}
  \label{eq:avl}
  \langle\ell\rangle = \sum_{\ell=1}^\infty \ell \
  \mathrm{Prob}(\mathrm{click},\ell) =
  \sum_{\ell=1}^\infty \ell p^2 (1-p)^{\ell-1} = \frac{2-p}{p},
\end{equation}
and the average $\ell^2$,
\begin{equation}
  \langle\ell^2\rangle = \sum_{\ell=1}^\infty \ell^2 \
  \mathrm{Prob}(\mathrm{click},\ell) =
  \sum_{\ell=1}^\infty \ell^2 p^2 (1-p)^{\ell-1} = \frac{p^2-6p+6}{p^2}.
\end{equation}
Thus the variance $\Var{\ell}$ is given by
\begin{equation}
  \label{eq:varl}
  \Var{\ell} = \langle\ell^2\rangle - \langle\ell\rangle^2 =
  \frac{2-2p}{p^2}.
\end{equation}

\begin{figure}
  \centering
  \includegraphics[width=0.75\textwidth]{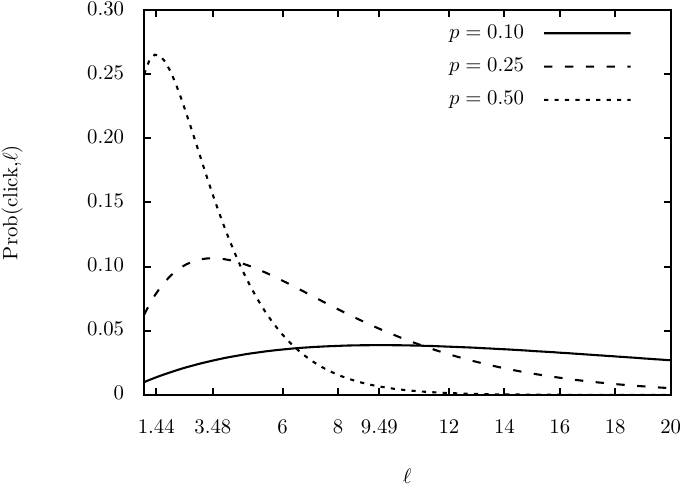}
  \caption{\label{fig1} The probability (\ref{eq:Prob}) of a~single
    click during $\ell$ steps of the clock's evolution. The values of
    $p$ are $0.1$, $0.25$, and $0.5$.}
\end{figure}

\begin{figure}
  \centering
  \includegraphics[width=0.75\textwidth]{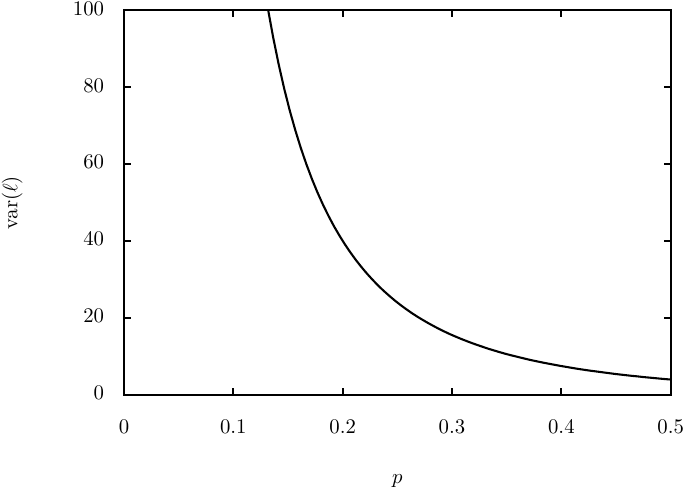}
  \caption{\label{fig2} The variance (\ref{eq:varl}) of the number of
    steps $\ell$ as a~function of $p$.}
\end{figure}

The probability Eq.~(\ref{eq:Prob}) is presented on Fig.~\ref{fig1} as
a~function of the number of steps $\ell$. We notice that for higher
probabilities $p$ the number of required evolution steps $\ell$
decreases. The exact position of the maxima is given by the relation
\begin{equation}
  \ell_{\mathrm{max}} = -\frac{1}{\ln(1-p)},
\end{equation}
which for $p=0.1$ gives $\ell_{\mathrm{max}}=9.49$, for $p=0.25$ is
$\ell_{\mathrm{max}}=3.48$, and for $p=0.5$ drops to
$\ell_{\mathrm{max}}=1.44$.

The variance of $\ell$ is presented on Fig.~\ref{fig2}. For the ideal
case of $p=0.5$, the average $\ell$ from Eq.~(\ref{eq:avl}) is
$\langle \ell\rangle = 3$ while the standard deviation reads
$\sigma = \sqrt{\mathrm{var}(\ell)} = 2$. This gives the average number
of the required evolution steps $3\pm 2$.

\section{Final remarks}

One may ask, how to compare the discussed quantum clock with the
currently used atomic clocks? The time unit measured by an atomic clock
is derived from the frequency of the photon emitted by an atom during
its de-excitation. From the perspective of our model, the stochastic
process of the atomic de-excitation plays the role of the proper clock,
while the interface can be found in the clock's system which detects the
photon.

The stability of the chosen atomic transitions is so high, that modern
atomic clocks \cite{zegar2013,zegar2015,zegar2016,zegar2021} achieve
systematic uncertainty on the level of $10^{-18}$s, where this number is
measured with respect to an external laboratory clock. Since each
evolution step advances the quantum clock on the time axis by $\xi$ and
we are required to wait between 1 and 5 steps to register the click, the
value of $\xi$ must be small enough to assure the already achieved
precission. If we allow $\xi$ to be a~random variable, with different
values in each step of the clock's evolution, the average
$\langle\xi\rangle$ should be small. In that case one may take the
statistics from as many as needed evolution steps, which will lower the
value of $\mathrm{var}(\ell)$ and increase the clock's accuracy to the
required level.

We notice also that every clock, even theoreticaly considered as
a~quantum system, is influenced by spacetime and physical
fields. Observed changes in the clock can give information about the
temporal structure of these objects. This is an open problem for future
investigations.

\bibliographystyle{plain}
\bibliography{zegarek-preprint}

\end{document}